\documentclass[submission,copyright,creativecommons]{eptcs}
\providecommand{\event}{HAS 2013} % Name of the event you are submitting to
\usepackage{breakurl}             % Not needed if you use pdflatex only.

\usepackage{color}
\usepackage{array}
\usepackage{wrapfig}
\usepackage{amsmath,amssymb}
\usepackage{graphicx}
\usepackage{faktor}
\usepackage[caption=false,font=footnotesize]{subfig}

\title{Monitoring with uncertainty}
\author{Ezio Bartocci
\institute{Faculty of Informatics\\
Vienna University of Technology\\
Vienna, Austria}
\email{ezio.bartocci@tuwien.ac.at}
\and
 Radu Grosu
\institute{Faculty of Informatics\\
Vienna University of Technology\\
Vienna, Austria}
\email{radu.grosu@tuwien.ac.at}
}
\def\titlerunning{Monitoring with Uncertainty}

\def\authorrunning{Ezio Bartocci, Radu Grosu}
\begin{document}
\maketitle

\begin{abstract}
We discuss the problem of runtime verification of an instrumented program that misses to emit and to monitor some events. These gaps can occur when a monitoring  overhead control mechanism is introduced to disable the monitor of an application with real-time constraints.  We show how to use statistical models  to learn the application behavior and to ``fill in" the introduced gaps. Finally, we present and discuss some techniques developed in the last three years to estimate the probability that a property of interest is violated in the presence of an incomplete trace. 
\end{abstract}

\section{Problem description}

Runtime verification (RV) (or \emph{monitoring})~\cite{Havelund2002a,Barringer2010a} is a well-established technique to 
check whether the current execution of a program satisfies
a property of interest. In the last years, RV has increasingly gained popularity 
in both  the formal verification and software engineering communities, because in 
most of the cases it reveals to be 
more practical than exhaustive verification techniques such as model checking~\cite{Clarke1982,Queille1982} and more versatile
than classical software testing. Formally, the RV problem is to decide whether an execution trace 
$\tau$ of a program $P$ satisfies a temporal logic specification $\varphi$. RV is then usually  
performed by translating the formula $\varphi$ into a deterministic finite state machine  (DFSM)
$M_{\varphi}$ and by instrumenting the program $P$ so that it emits the events triggering 
 the input-enabled transitions in $M_{\varphi}$.

However, RV  does not come for free. It introduces runtime overhead, thereby
changing the timing-related behavior of the program under scrutiny. While this is acceptable
in many applications, it may be unacceptable in applications with real-time
constraints. In such cases, overhead control is necessary.
Recently, a number of techniques have been developed to mitigate the overhead
due to RV~\cite{Bonakdarpour2011,fei06artemis,hauswirth04leak,arnold08qvm,Huang2012}. Common to these approaches is the use of event
sampling to reduce overhead. Sampling means that some events are not processed
at all, or are processed in a limited (and thus less expensive) manner than other
events. In a previous work~\cite{Huang2012}, we introduced Software Monitoring with Controllable Overhead
(SMCO), an overhead-control technique that selectively turns monitoring on and
off, such that the use of a short- or long-term overhead budget is maximized and never
exceeded. Gaps in monitoring, however, introduce uncertainty in the monitoring results.

For example, let $\varphi$ be the formula $\square (a \Rightarrow \diamond c)$, that means
always an event $a$ is finally followed by an event $c$ and let $\tau$ be the trace $a$ $b$ $b$ $c$ $ a$ $d$ $b$ $c$. 
In this example the formula $\varphi$ clearly holds. Suppose now that the trace is
\emph{incomplete}, due to disabled monitoring: $\tau =$ $a$ $b$ $b$ $c$  $-$ $b$ $c$, 
with $-$ indicating a set of events that could not  be observed (or gap). Although 
it is not possible to be really sure that the property is satisfied, the problem we are 
interested to solve is to quantify the uncertainty with which the trace $\tau$ satisfies $\varphi$.

\section{``Filling in" the gaps}

In order to quantify the uncertainty in monitoring, our approach
is to use a statistical model of the monitored system
to ``fill in" sampling-induced gaps in event sequences,
and then calculate the probability that the property
of interest is satisfied or violated.
In our previous works~\cite{Stoller2011,Bartocci2012,Kalajdzic2013}, we have chosen 
the Dynamic Bayesian Networks (DBNs)~\cite{russellnorvig}, a suitable formalism to
characterize the temporal probability model of
the instrumented program emitting events. DBNs can
have multiple state variables (modeling the states
of a program that are usually hidden) and multiple
observation variables (representing the output events).
A DBN is essentially a first-order Markov Process
where each variable at time $t$ can depend only
by other variables at the same time $t$ or at time $t-1$.

In~\cite{Stoller2011,Bartocci2012} we used Hidden Markov 
Models (HMM),  that are essentially DBNs with only one state 
variable and one observation variable. A HMM is a Markov model in 
which the system being modeled is assumed 
to be a Markov process with unobserved (hidden) states. In a regular Markov 
model, states are directly visible to the observer, and therefore state transition 
probabilities are the only required parameters. In a HMM, states cannot be 
observed; rather, each state has a probability distribution for the possible 
observations (formally called observation symbols). The classic state estimation 
problem for HMMs is to compute the most likely sequence of states that generates 
a given observation sequence.

One can obtain a HMM for a system automatically, by
learning it from complete traces using standard HMM learning algorithms(i.e. Baum--Welch)~\cite{rabiner89tutorial}.
These algorithms require the user to specify the desired number of states in
the HMM and they allow (but do not require) the user to provide
information about the structure of the HMM, specifically, that certain entries
in the transition probability matrix and the observation probability matrix are
close to zero. This information can help the learning algorithm converge more quickly
and find globally (instead of locally) optimal solutions.

In contrast to our previous work~\cite{Stoller2011,Bartocci2012}, 
in~\cite{Kalajdzic2013} we succinctly represent the program model, the program
monitor, their interaction, and their observations as a generic DBN.
This allowed us to properly formalize a new kind of event, called \emph{peek events}, 
which are inexpensive observations of part of the program state. In many applications, program
states and monitor states are correlated, and hence peek events can be used to narrow
down the possible states of the monitor DFSM. We use peek events at the end of monitoring
gaps to refocus the DBN and DFSM states. Our combination of these two kind of
observations, program events and peek events, is akin to \emph{sensor fusion} in robotics.

\section{Monitoring with uncertainty}

\subsection*{Runtime Verification with State Estimation (RVSE)}
To quantify the uncertainty, one can estimate the current state of the program.
We developed a framework for this, called Runtime Verification with State Estimation
(RVSE)~\cite{Stoller2011}, in which a HMM is used to succinctly model the
program and the uncertainty in predictions due to incomplete information.

While monitoring is on, the observed program events drive the transitions of the
property checker, modeled as a deterministic finite state machine (DFSM). They also provide
information used to help correct the state estimates (specifically, state probability
distributions) computed from the HMM transition probabilities, by comparing the output
probabilities in each state with the observed outputs. When monitoring is off, the
transition probabilities in the HMM alone determine the updated state estimate after the
gap, and the output probabilities in the HMM drive the transitions of the DFSM. Each
gap is characterized by a gap length distribution, which is a probability distribution for
the number of missed observations during that gap.
Our algorithm was based on an optimal state estimation algorithm, known as the
forward algorithm, extended to handle gaps. Unfortunately, this algorithm incurs high
overhead, especially for longer sequences of gaps, because it involves repeated matrix
multiplications using the observation-probability and transition-probability matrices. In
our measurements, this was often more than a factor of 10 larger than the overhead of
monitoring the events themselves.
\subsection*{Approximate Precomputed Runtime Verification with State Estimation (AP-RVSE)}
To reduce the runtime overhead, we developed a version of the algorithm, which we
call the approximate precomputed RVSE (AP-RVSE), which pre-computes the matrix
calculations and stores the results in a table~\cite{Bartocci2012}. 
Essentially, AP-RVSE pre-computes a potentially infinite graph unfolding, where nodes 
are labeled with state probability distributions, and edges are labeled with transitions. 
To ensure the table is finite, we introduced an approximation in the calculations, controlled by an accuracy
$\epsilon$ parameter: if a newly computed matrix differs from the matrix on an existing node by at most
$\epsilon$ according to the 1-norm, then we re-use the existing node instead of creating a new one.
With this algorithm, the runtime overhead is low, independent of the desired accuracy,
but higher accuracy requires larger tables, and the memory requirements could become
problematic. Also, if the set of gap length distributions that may appear in an execution
is not known in advance, precomputation is infeasible.

\subsection*{Runtime Verification with Particle Filtering (RVPF)}

In a recent paper~\cite{Kalajdzic2013} we have introduced an alternative approach, 
called Runtime Verification with Particle Filtering (RVPF), to control the balance between 
runtime overhead, memory consumption, and prediction accuracy. In one of the most 
common forms of particle filtering (PF)~\cite{Gordon1993}, the probability distribution of states 
is approximated by the proportion of particles in each state. The particle filtering process works 
in three recurring steps. First,
the particles are advanced to their successor states by sampling from the HMM's transition
probability distribution. Second, each particle is assigned a weight corresponding
to the output probability of the observed program event. Third, the particles are resampled
according to the normalized weights from the second step; this has the effect of
redistributing the particles to the states to provide a better prediction of the program
events. We exploit the knowledge of the current program event and the particular structure
of the DBN to improve the variance of the PF, by using sequential importance resampling
(SIR). In this PF variation, resampling (which is a major performance bottleneck)
does not have to be performed in each round, and the particles are advanced to their
successor states by sampling from the HMM's transition probability distribution
conditioned by the current observation. 
%While this conditional probability distribution cannot be computed in general, it can be computed for HMMs.

Adjusting the number of particles used by RVPF provides a versatile way to tune the
memory requirements, runtime overhead, and prediction accuracy.With larger numbers
of gaps, the particles get more widely dispersed in the state space, and more particles are
needed to cover all of the interesting states. To evaluate the performance and accuracy of
RVPF, we implemented it in~\cite{Kalajdzic2013}  along with our previous two algorithms
 in C and compared them through experiments based on the benchmarks used in~\cite{Bartocci2012}. 
Our results confirm RVPF's versatility.

\paragraph{Acknowledgements.}

The main concepts and ideas presented in this invited talk were developed (and presented in several papers) 
 in collaboration with Scott A. Smolka, Scott D. Stoller, Erez Zadok, Justin Seyster 
and Klaus Havelund that we would like here to acknowledge.

\nocite{*}
\bibliographystyle{eptcs}
\bibliography{bibtex}
\end{document}